%% file: 2015_phantom.tex
\documentclass{elsarticle}

\usepackage{tabularx}
\makeatletter
\newif\if@restonecol
\makeatother

\usepackage{algorithm2e}

\usepackage{epsfig}
\usepackage{graphicx}
\usepackage{subfloat}
\usepackage{caption}
\usepackage{amsmath, amsthm, amssymb}
\usepackage{subcaption}
\usepackage{color}
\usepackage{url}
\usepackage{comment}
\usepackage[font={small,bf},labelfont=bf]{caption}
\DeclareCaptionType{copyrightbox}

\newtheorem{definition}{Definition}

\newcommand{\un}{phantom}
\newcommand{\para}[1]{{\vspace{4pt} \bf \noindent #1}}

\begin{document}
\begin{frontmatter}
\author{V{\'a}clav Bel{\'a}k}
\address{Insight Centre for Data Analytics\\Galway, Ireland}
\ead{vaclav@belak.net}
\author{Afra Mashhadi}
\address{Bell Labs\\Dublin, Ireland}
\ead{afra.mashhadi@alcatel-lucent.com}
\author{Alessandra Sala}
\address{Bell Labs\\Dublin, Ireland}
\ead{alessandra.sala@alcatel-lucent.com}
\author{Donn Morrison}
\address{Department of Computer and Information Science\\Norwegian University of Science and Technology\\Trondheim, Norway}
\ead{donn.morrison@deri.org}

\title{Phantom cascades: The effect of hidden nodes on information diffusion}

\begin{abstract}
Research on information diffusion generally assumes complete knowledge of the underlying network. However, in the presence of factors such as increasing privacy awareness, restrictions on application programming interfaces (APIs) and sampling strategies, this assumption rarely holds in the real world which in turn leads to an underestimation of the size of information cascades. In this work we study the effect of hidden network structure on information diffusion processes. We characterise information cascades through activation paths traversing visible and hidden parts of the network. We quantify diffusion estimation error while varying the amount of hidden structure in five empirical and synthetic network datasets and demonstrate the effect of topological properties on this error. Finally, we suggest practical recommendations for practitioners and propose a model to predict the cascade size with minimal information regarding the underlying network.

\end{abstract}

\end{frontmatter}



\input{inc/intro}

\input{inc/related}

\input{inc/methodology}

\input{inc/results_emprical2_ale}

\input{inc/results_theo}

\section{Discussion}
\input{inc/implication}

\input{inc/conclusion}

\section*{Acknowledgments}

This work was funded by Science Foundation Ireland (SFI) under the CLIQUE (SFI/08/SRC/I1407) and Insight (SFI/12/RC/2289) research grants.

{
\section*{References}
\bibliographystyle{abbrv}
\bibliography{references}  
}
\end{document}

%% file: inc/intro.tex
\section{Introduction}\label{intro}

Simulating information diffusion processes is a critical aspect of understanding how information spreads in real world networks. A word-of-mouth viral marketing campaign is a well known example where a company wishes to estimate the spread of an advertisement or uptake of a product over a social network.
A prerequisite for such studies is access to real network data over which the processes of diffusion can be studied. Online social networks (OSNs) are a natural choice for this purpose as they are readily available and in many cases constitute the desired diffusion medium (e.g., Facebook and Twitter).  Consequently such datasets have been widely used in previous research~\cite{yang2010modeling,DBLP:conf/socialcom/DuongWS11,sadikov2011correcting,myers2012}.

\begin{figure}
\centering
\includegraphics[width=\textwidth]{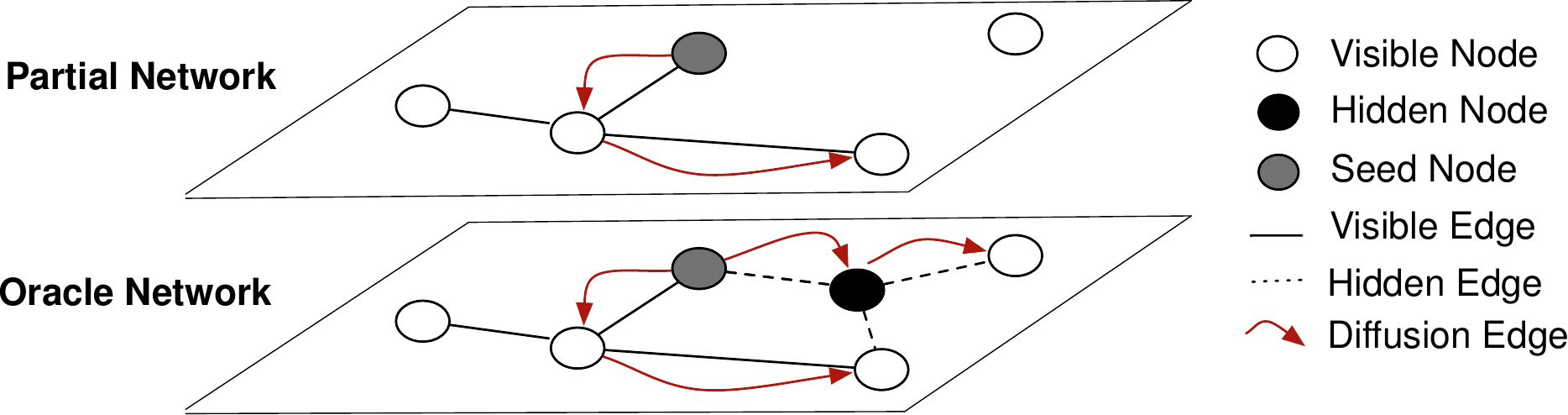}

\caption{\small An information cascade on an oracle (complete) network and its partially observed counterpart.}
\label{fig_partnet}
\vspace{-1em}
\end{figure}

However, access to the complete network in question is rarely possible. Factors such as privacy settings, application programming interface (API) restrictions and sampling strategies contribute to missing network structure. For example, it is well known that users of OSNs are growing increasingly privacy-aware. A recent large-scale study of 1.4 million Facebook users~\cite{DBLP:conf/percom/DeyJR12} revealed an increase in privacy-enabled profiles over 15 months from 17\% to more than 50\%, effectively rendering those users hidden from study yet still actively connected to their friends. A possible implication of this trend is that as these networks become more and more \textit{partial}, research using this data to simulate the spread of information may become less accurate. The problem is illustrated by the example in Figure~\ref{fig_partnet} showing a hypothetical information cascade over two networks: a complete network referred to as the \emph{oracle}, where information about all nodes and edges is known, and a partial view, where some portion is hidden. 

This leads to an important question: how does partial network knowledge affect models of information diffusion? In other words, \emph{how can we quantify the error introduced as we move away from the completeness assumption of the underlying network?, and how can we correct for it?} These are the central motivating questions we aim to address in this work. We focus on the \textit{Independent Cascade Model} (ICM) \cite{kempe2003maximizing,leskovec2007cost,chen2009efficient,myers2012} as a classic example of an information diffusion algorithm and measure the differences between information cascades on oracle and partial networks. To this end we formulate a three-step approach that begins with a complete network, samples a set of nodes of increasing size to be marked \textit{hidden} (i.e., modelling users who have privacy enabled), and finally simulates information spreading across the resulting networks.

With the proposed methodology we are able to quantify the error introduced due to network partiality based on a theoretical oracle scenario (i.e., how \textit{would have} the information spread with full knowledge). We pay particular attention to the paths that lead to the activation of visible nodes in the oracle network by distinguishing whether or not they were activated via hidden nodes. We refer to this class of activated nodes as the \emph{{\un } cascade} and set out to understand how the size of this cascade changes as a function of the percentage of hidden nodes. 

 As we explain in Section~\ref{sec:related}, there is limited prior work studying the effect of partial information on diffusion models~\cite{Cebrian:2009:WAY:1641389.1641391,sadikov2011correcting,DBLP:journals/corr/abs-1210-3587} from which our work differs in two fundamental aspects: i) by distinguishing between cascade types we are able to compare the observed cascade on the partial network to a theoretical cascade on the oracle network, ii) we focus on missing data at the network level (as opposed to the cascade level~\cite{sadikov2011correcting,DBLP:journals/corr/abs-1210-3587}) and study how information cascades over the incomplete network. To the best of our knowledge, this is the first study to quantify the error related to the diffusion cascade caused by nodes hidden at the network level.

The remainder of this paper is organised as follows. In Section \ref{sec:related} we discuss related work on missing data in information cascades. In Section \ref{sec:methodology} we present the datasets and the methodology we use to measure the effect of partial knowledge on the diffusion model. Section \ref{sec:experimental_result} presents the results in which we demonstrate the magnitude of the problem and its relation to topological properties specific to different networks. In Section \ref{sec:prediction} we introduce a model correcting for the effect of missing data. In Section \ref{sec:implications} we discuss the implication and limitations of this work.  Finally we envisage  our future directions  and conclude this study in Section~\ref{sec:conclusion}.

%% file: inc/related.tex
\section{Related work}
\label{sec:related}

The problem of missing data in networks has been addressed from different perspectives ranging from network sampling \cite{rothenberg1995sampling, citeulike:223029, gjoka2010walking}, where the aim is to obtain a representative subset of the network, to network reconstruction \cite{kim2011network, DBLP:conf/socialcom/DuongWS11}, where nodes and edges are inferred to recreate the original network. Other works, such as \cite{citeulike:223029,kossinets2006effects}, have examined the effect of missing nodes and edges on topological metrics of the network (e.g., average node degree, diameter, clustering coefficient). These and other studies~\cite{Najar:2012:PID:2187980.2188261, Lin:2012} have helped to build an understanding of the implications of research conducted on sampled or otherwise incomplete network data.

In the context of information diffusion, however, very little research has been conducted and thus the effect of missing data on the information cascades themselves is not well understood. Of the vast amount of research on information spreading and diffusion, we are aware of only four studies partially addressing this problem~\cite{Cebrian:2009:WAY:1641389.1641391, sadikov2011correcting, DBLP:journals/corr/abs-1210-3587, yang2010modeling}. In~\cite{Cebrian:2009:WAY:1641389.1641391}, the authors uncovered a logarithmic error as a function of the amount of missing data for diffusion simulations on a small telecommunications call graph. Diffusions are simulated and compared on both an oracle and partial graph. However, the partial graph is created by removing nodes, not hiding them, so the role they play in the oracle is not studied. On the other hand, our approach permits comparison between theoretical spread on the oracle and observed spread on the partial network through a characterisation of cascades based on activation paths, yielding insight into how different cascades contribute to diffusion error.

The other three approaches attempt to infer properties of the total information cascade from a partially observed cascade~\cite{sadikov2011correcting, DBLP:journals/corr/abs-1210-3587,yang2010modeling}. In~\cite{sadikov2011correcting}, the authors address cascade distortions under missing data and propose fitting $k$-trees to correct the distortion. In~\cite{DBLP:journals/corr/abs-1210-3587} a similar approach is taken but the authors incorporate node activation time to constrain the fit of \textit{consistent} trees to the information cascades, while~\cite{yang2010modeling} exploits node activation time in a model that predicts the final number of activated nodes without knowledge of the network. The key differences between these works and ours are i) they focus on missing data at the cascade level, whereas we focus on missing data at the network level, ii) they examine cascades that have already occurred, while we aim to estimate the error under missing data prior to an attempted real world diffusion (e.g., a viral marketing campaign) and propose a model to correct for it. Furthermore, in~\cite{DBLP:journals/corr/abs-1210-3587}, the authors assume \textit{complete} knowledge of the underlying network in order to fit the cascade trees, which is rarely the case in the real world.

%% file: inc/methodology.tex
\section{Methodology}\label{methodology}
\label{sec:methodology}

In this section we aim to study the effect of partial network knowledge on the diffusion process using five datasets representing both empirical and synthetic social networks. We describe the creation of partial networks and propose a novel methodology to quantify the error due to missing information by characterising the different paths through which a node can become activated during the diffusion process.

\begin{table}[t]
  \begin{center}
  \begin{tabular}{lcccc}
    \hline
    Network                      & \# Nodes   & \# Edges  & Nature  \\
    \hline
    DBLP                         & 1103412 & 4225686 &  Co-authorship 	\\ 
    Astro Physics                & 18772   & 198110  & Co-authorship \\ 
    2.5K series (AstroPh)  & 16619   & 158218  &  Social Network \\ 
    TOSHK                        & 100000  & 1062105 &  Social Network \\ 
    Erd\H{o}s-R\'{e}nyi          & 100000  & 1048894 &  Random \\
    \hline
  \end{tabular}
  \end{center}
    \caption{\small Empirical (DBLP,  Astro Physics) and synthetic (TOSHK, Erd\H{o}s-R\'{e}nyi, 2.5K series) networks.}
  \label{tab:networks1}

  \small

\end{table}

\subsection{Datasets}
Three factors drove the choices in datasets. First, our aim is to use graphs that are arguably as \textit{complete} as possible to serve as \textit{oracle} networks, i.e., where all nodes and edges are known. We refer to them as oracle networks because they represent \textit{relatively} complete (in the case of empirical) or theoretically complete (in the case of synthetic) social networks.\footnote{The notion of completeness could indeed be argued for a known sampled network, given that it could play the role of oracle compared to a partial network derived by sampling it. However, for simplicity we define a complete network as one that is, to the best of our knowledge, complete in that it encompasses all possible nodes and edges (i.e., it is not sampled from a larger network).} Secondly, to account for the effect of network size we chose networks of different scale. Finally, the graphs should have properties consistent with social networks (i.e. have community structure, exponential degree distribution, small diameter, etc.) because we are simulating a process from the social network domain, that is, where a message is spread by word-of-mouth.

\para{Empirical.} We selected two empirical  networks: the DBLP co-authorship network
\cite{DBLP:conf/spire/Ley02} and the ArXiv Astro Physics co-authorship network \cite{b242}. These networks were chosen because
they represent the flow of knowledge through scientific collaboration and are examples of relatively complete networks
for those communities.\footnote{The DBLP network is a multigraph with parallel edges between authors
  denoting multiple co-authored papers. However, in this work we ignore parallel edges in order to leverage the seed selection strategy
  proposed by~\cite{chen2009efficient} and to maintain consistency across datasets.}

\para{Synthetic.} In addition to the empirical datasets, we generated random networks with different topological properties. Generating synthetic graphs has the benefit of yielding networks of desired size and topological properties, providing a benchmark comparison for the real networks and a more complete picture of the results. Furthermore, the resultant graphs are by definition complete, as they are realised from a generative process. We selected three graph models: 2.5K series \cite{gjoka13_2.5K_Graphs}, TOSHK \cite{Toivonen2006851}, and the Erd\H{o}s-R\'{e}nyi random graph \cite{Erdos60onthe}. 

The 2.5K series requires a graph as input to explicitly model the joint degree distribution (JDD) and degree-dependent average clustering coefficient (DACC). The method works by replicating the JDD, overestimating the number of triangles, and then systematically breaking triangles with double edge swaps until the target DACC is reached. Due to the complexity of this method, we fit our smallest empirical network, i.e., Astro Physics, by following the methodology presented in \cite{gjoka13_2.5K_Graphs}. Note that the exact number of nodes cannot be targeted which is the reason this property does not exactly match the target network. 

TOSHK is a social network growing model that targets two properties: an exponential degree distribution and high clustering coefficient. We tune these features based on the Astro Physics network (for consistency with 2.5K series) using a grid search resulting in parameters $k=22$ (average degree) and $p=0.9$ (probability of connecting to a given node's neighbours), however we increase the size of the network to $N=100000$ (nodes) to explore different size effects. 

Finally, the Erd\H{o}s-R\'{e}nyi random graph was included for comparison (not having any social structure) and was generated with $N=100000$ (nodes) and $p=0.00021$ (edge probability) to match the average degree of Astro Physics. Table~\ref{tab:networks1} presents an overview of the described networks.

\subsection{Partial network creation}

To simulate an incomplete OSN crawled or otherwise collected we create a partial network from the oracle by uniformly sampling nodes and edges for removal. The decision to sample uniformly is motivated by evidence that users are not influenced by their friends in deciding to be more private~\cite{DBLP:conf/percom/DeyJR12}. In addition, some OSNs, such as Twitter, uniformly sample information to be displayed to users \cite{sadikov2011correcting}.\footnote{In the event that correlation would exist between hidden nodes of a given network, uniform sampling can be readily replaced by a more appropriate strategy (random walk, etc.).} We chose the node removal interval to be 10--50\% in 10\% increments to model privacy rates empirically measured \cite{DBLP:conf/percom/DeyJR12} and make the assumption that missing data in other OSNs would likely fall within this range.

Formally, we define an {\em oracle} graph as $G_o=(V_o,E_o)$ where $V_o$ and $E_o$ represent the nodes and undirected edges in the oracle network. To create a {\em partial} view $G_p$ of $G_o$, we uniformly sample a fraction $\rho$ of {\em hidden} nodes $V_h$ from $V_o$ where $\rho \in$ $\{0.1,0.2,0.3,0.4,0.5\}$. Once a node is selected by the sampling strategy as hidden, all its edges are also hidden from the partial network. The resulting partial graph is thus defined as $G_p=(V_p,E_p)$ where $V_p=V_o\setminus V_h$ and $E_p=\{(u,v) \in E_o~ | ~ u,v \in V_p \}$.

\subsection{Independent Cascade Model}

The Independent Cascade Model (ICM) is a well-known model for the spread of information over social networks~\cite{kempe2003maximizing}. We focus solely on this model because it has been extensively studied for various diffusion problems such as seed selection~\cite{kempe2003maximizing,chen2009efficient} and spread maximisation~\cite{leskovec2007cost,myers2012} and thus we can take advantage of previously suggested guidelines and parameter settings. We will discuss the possible impacts  of these choices  in Section~\ref{sec:implications}.

Given a graph $G=(V,E)$, a set of seed nodes $S \subseteq V$ initially actived at time $t=0$ and all other nodes $V \setminus S$ initially inactive, ICM works in a push mode to spread information over the edges $E$. At each iteration $t\ge1$, every node $i$ that was activated at time $t-1$ has exactly one chance to activate each of its inactive neighbours $j$ with transmission probability $p$. ICM terminates when no further nodes are activated.

\para{Transmission probability.} Following previous work~\cite{chen2009efficient,kempe2003maximizing}, we assume a global transmission probability $p$ for all nodes in $G$. Previous empirical work has shown evidence supporting low global transmission probabilities. In \cite{Bakshy:2012:RSN:2187836.2187907}, for example, the authors find a 0.191\% likelihood of resharing friends' posts on Facebook. While it is difficult to draw a conclusion from this on an ideal value for $p$, we report results using $p=0.01$, but note that the same experiments were conducted with $p=0.001$ and we observed consistent results. We did not consider a higher transmission probability such as $p=0.1$ because it is known to quickly saturate the network~\cite{chen2009efficient}.

\para{Seed size and selection.} The number of seeds $|S|$ and the heuristic to select them is another important parameter to ICM which has attracted a large body of research, among which extensive analyses can be found in~\cite{leskovec2007cost,chen2009efficient}. For seed selection we chose the \emph{discounted degree} strategy~\cite{chen2009efficient}, which selects an additional seed node by its degree adjusted by the estimated number of nodes, that will be activated by already chosen seeds. This heuristic has been shown to have a better performance than plain degree, and it guarantees the seed selection from the core of the network. 

Most importantly, the resulting cascades from the oracle and partial network have to be comparable. In order to ensure this we use the same seeds in both scenarios. To this end we first apply the seed selection algorithm on the {\emph partial} network and maintain this set for the oracle network. Finally, as each dataset differs significantly in size, we select the number of seeds to be a fraction $\gamma$ of the number of nodes in the partial network ($V_p$) where $\gamma\in\{0.0001,0.001,0.01\}$.

\subsection{Experimental setup}

We run ICM in two scenarios:
\begin{enumerate}
\item the oracle scenario where ICM is run over the complete network allowing the information to spread over the hidden nodes and edges, and
\item the partial scenario in which information spreads only over the visible part of the oracle network (the partial network).
\end{enumerate}

\begin{figure}[!t]
\centering
\includegraphics[width=.7\textwidth]{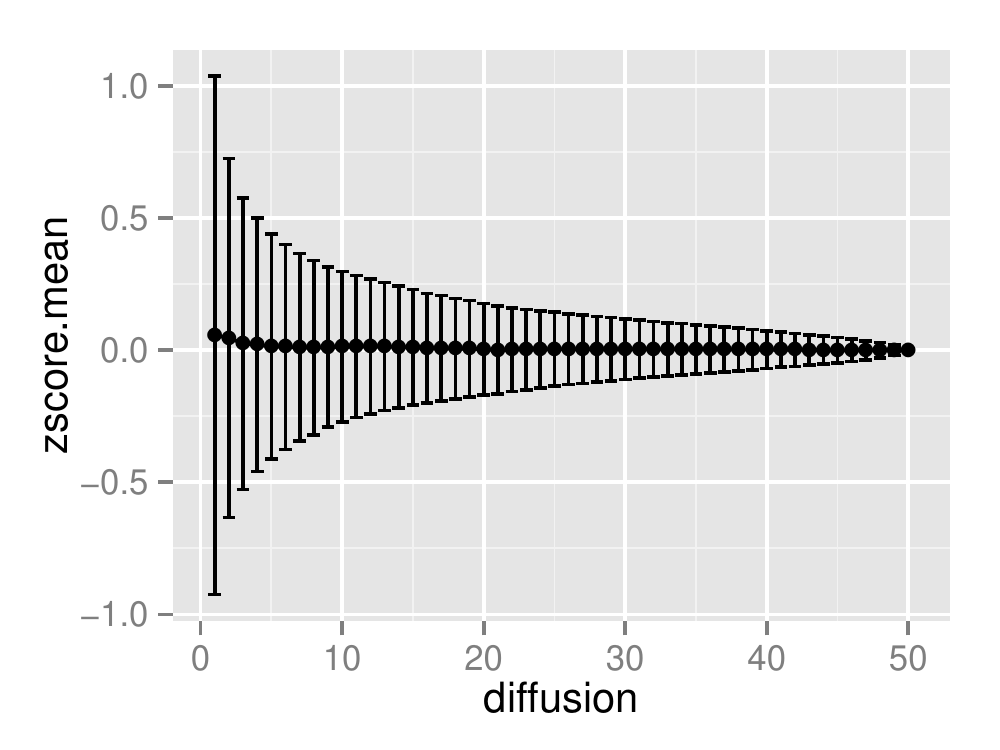}

\caption{\small Mean z-score (y-axis) of the z-scores measuring the difference between the simulated cascade size after $r$ diffusion trials (x-axis) and the final mean obtained at $r=50$. 
The error bars depict one standard deviation from the mean z-score.}
\label{fig:convergence}

\end{figure}

For each scenario, we conducted experiments for each fraction of hidden nodes $\rho \in$ $\{0.1,0.2,0.3,0.4,0.5\}$.
We first account for variation due to different samplings of hidden nodes by generating a total of $v$ partial views (i.e., samples) from the oracle network.
Next, to account for variance due to ICM across each sample, we run each diffusion $r$-times for each {\em sample} of the oracle and partial networks.
We empirically determined the minimum required values for both $r$ and $v$ to be $50$, that is 50 samples per experiment and 50 diffusions per sample. 
As can be seen in Figure~\ref{fig:convergence}, the mean cascade size quickly converged and thus this number of repetitions allows us to place an upper bound on the number of required experiments while maintaining diffusions
that are computationally tractable.
\footnote{In the case of the largest dataset, DBLP, all experiments took 29 hours to compute using 24 Intel Xeon 2.4 GHz CPUs.}

Due to the stochastic nature of ICM, it is not possible to directly compare a single diffusion on the oracle network with a single diffusion on the partial network. However, it is perfectly valid to compare averages from the multiple diffusions over the different samples. 

\begin{figure}[!t]
\centering
\includegraphics[width=.7\textwidth]{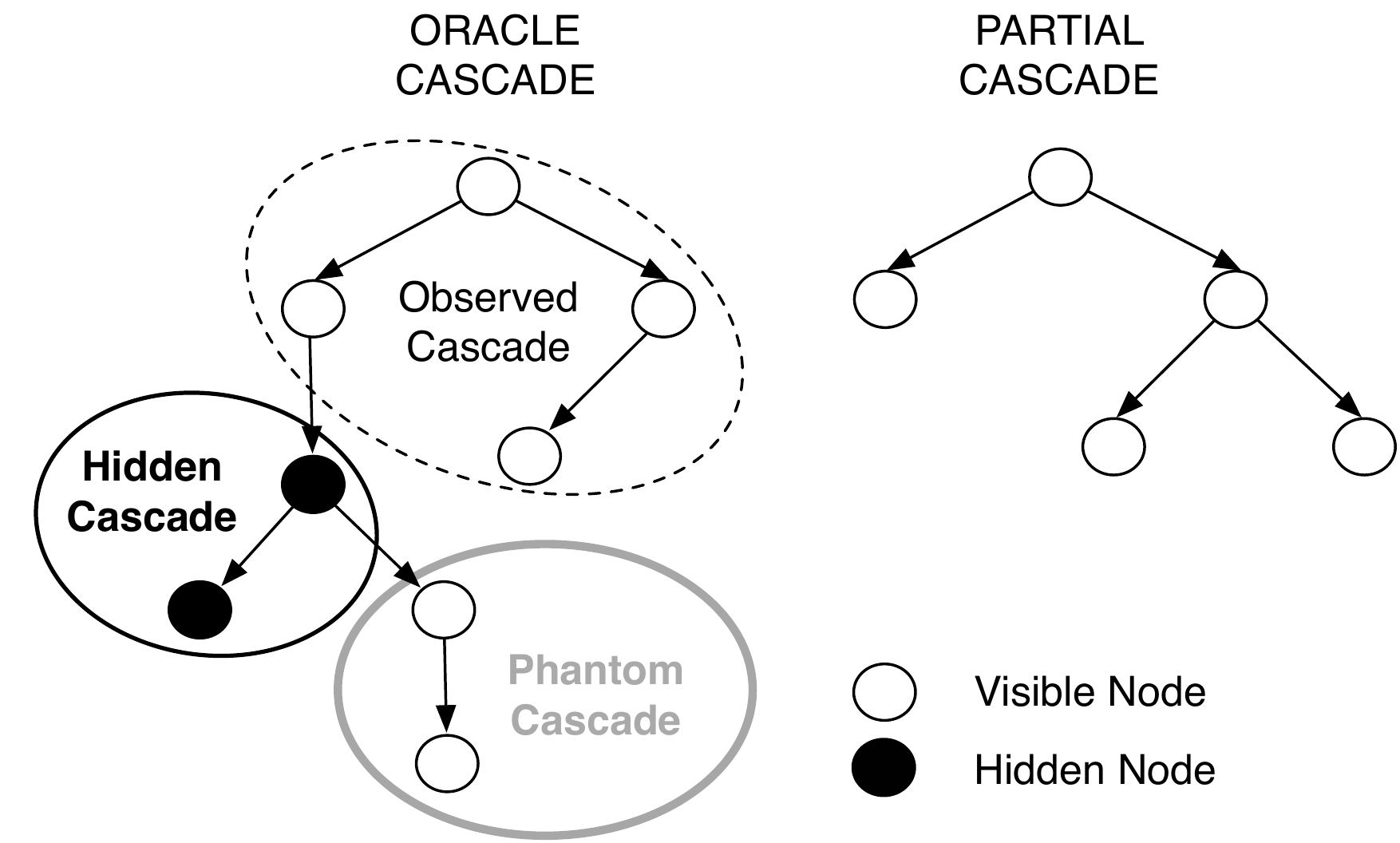}

\caption{\small Information cascades for the oracle (left) and partial (right) networks.}
\label{fig_cascade_defns}

\end{figure}

\subsubsection{Characterisation of activation paths and cascade sizes}
Each diffusion run results in a set of information cascades with the seeds as the root nodes. Let us first focus on the diffusion process in the oracle network and the cascades that are formed as a result. Consider an information cascade $C_s$ represented as a tree with seed node $s$ as its root and where an edge $(i,j)$ represents the activation of node $j$ by node $i$. 

We define three distinct ways that an arbitrary node $j$ can be activated in the oracle network and illustrate the resulting cascades in Figure~\ref{fig_cascade_defns}.
\vspace{0.1cm}
\begin{definition}
A node $j$ is \emph{observably activated} if $j$ is a visible node ($j\in V_p$) and the activation path to $j$ does not contain any hidden nodes.
\end{definition}
\begin{definition}
A node $j$ is \emph{phantomly activated} if $j$ is a visible node ($j\in V_p$) and the activation path contains at least one hidden node.
\end{definition}
\begin{definition}
A node $j$ is \emph{hiddenly activated} if $j$ is a hidden node ($j \in V_h$) regardless of whether it was activated through a visible or a hidden path.
\end{definition}

Finally as in ICM, a node can be successfully activated by multiple neighbours, we compute the expected value of phantomly activated and observably activated nodes by considering the probability of activation by multiple parents, some of which may be visible, while the others may be phantom or hidden. For example, if a visible node is activated by two nodes, one visible and one hidden, it will contribute by 0.5 to the $\sigma_o$ and by 0.5 to $\sigma_{ph}$.

We now turn our attention to measuring the sizes of the information cascades.
\begin{definition}
The \emph{total cascade size}, denoted by $\sigma$, is defined as the total number of nodes activated during the diffusion on the oracle network and corresponds to the total number of nodes in the cascade.
\end{definition}
\begin{definition}
The \emph{observed cascade size}, denoted by $\sigma_o$, is defined as the number of observably activated nodes in the oracle network. We similarly define and denote the \emph{{\un } cascade size} and \emph{hidden cascade size} as $\sigma_{ph}$ and  $\sigma_h$ respectively.
\end{definition}

As these cascades are each a subset of the diffusion tree, their sum corresponds to the total cascade size, that is $\sigma=\sigma_o+\sigma_{ph}+\sigma_h$.

Finally the resulting cascades from the diffusion process on the partial network are defined as:

\begin{definition}
The \emph{partial cascade size}, denoted by $\sigma_p$, is defined as the total number of nodes activated during the diffusion on the partial network. 
\end{definition}

\subsubsection{Metrics}
\label{sec:spreadresid}

Measuring the partial cascade size allows us to quantify the measurement error caused by partial knowledge of the network. We measure this error in terms of relative error of the activated nodes, while accounting for the cascades traversing hidden nodes unlike the previous approaches~\cite{Cebrian:2009:WAY:1641389.1641391}. Distinguishing between {\un } and hidden cascades allows us to account for what theoretically would have happened to the same visible nodes in the partial network if knowledge of the oracle network was available.

\begin{equation}
  \label{eq:residual}
\mbox{\textit{relative error}} =\frac{{\sum_{i\in v} \frac{|((E(\sigma_{ph})+E(\sigma_o))- E(\sigma_p)|}{(E(\sigma_{ph})+E(\sigma_o))}}}{v},
\end{equation}

\noindent where $E(\sigma_\cdot)$ is the expected cascade size over the $r$ diffusions on each sample, and $v$ is the size of the sample set.
The relative error is a simple metric to capture the error in the partial cascade estimation by taking into account the size of the phantom and observed cascades.

Note that we do not focus on the size of the hidden cascade $\sigma_h$ in the estimation of our error. Not only will including $\sigma_h$ simply increase the relative error, but, from an application perspective, the hidden nodes may be of less interest to the campaigners  than those visible nodes which are the focus of the campaign (i.e., diffusion process).  Take the example of a telecommunications operator who wishes to promote a new service to existing customers. Only the operator's own customers are of interest, but possible diffusion paths to existing clients can exist that include parts of other provider's networks (i.e., through network structure hidden to the operator).

%% file: inc/results_emprical2_ale.tex
\section{Quantifying diffusion error}
\label{sec:experimental_result}

This section presents an overview of the results of the experiments  detailed in the previous section. We first examine the problem through a sample use case, highlighting the impact of hidden nodes on what is observed from the diffusion process. We then investigate this problem in depth by quantifying the diffusion error, measured by relative error, under partial knowledge. Finally, we demonstrate how the fraction of hidden nodes, the seed size and the network topology contribute to the observed effects.

\subsection{Use case}

To illustrate the impact of incomplete network data on the observed diffusion process, consider the following hypothetical scenario. A company is launching an advertising campaign on a social network where an initial group of users (the seeds) will be given an incentive to spread to friends via word-of-mouth. To realise such scenario, let us consider the 2.5K network which has been shown to incorporate the typical social network structure~\cite{gjoka13_2.5K_Graphs}. Assume that 50\% of the users of this social network ($N=16619$) are privacy aware and restrict access to profiles, resulting in a partial view of the network that is collected by the company. Furthermore, it has been decided that for the campaign to be considered successful (e.g., to warrant the investment), it must reach least 100 users.

\begin{figure}[t]
  \begin{center}

\includegraphics[width=\textwidth]{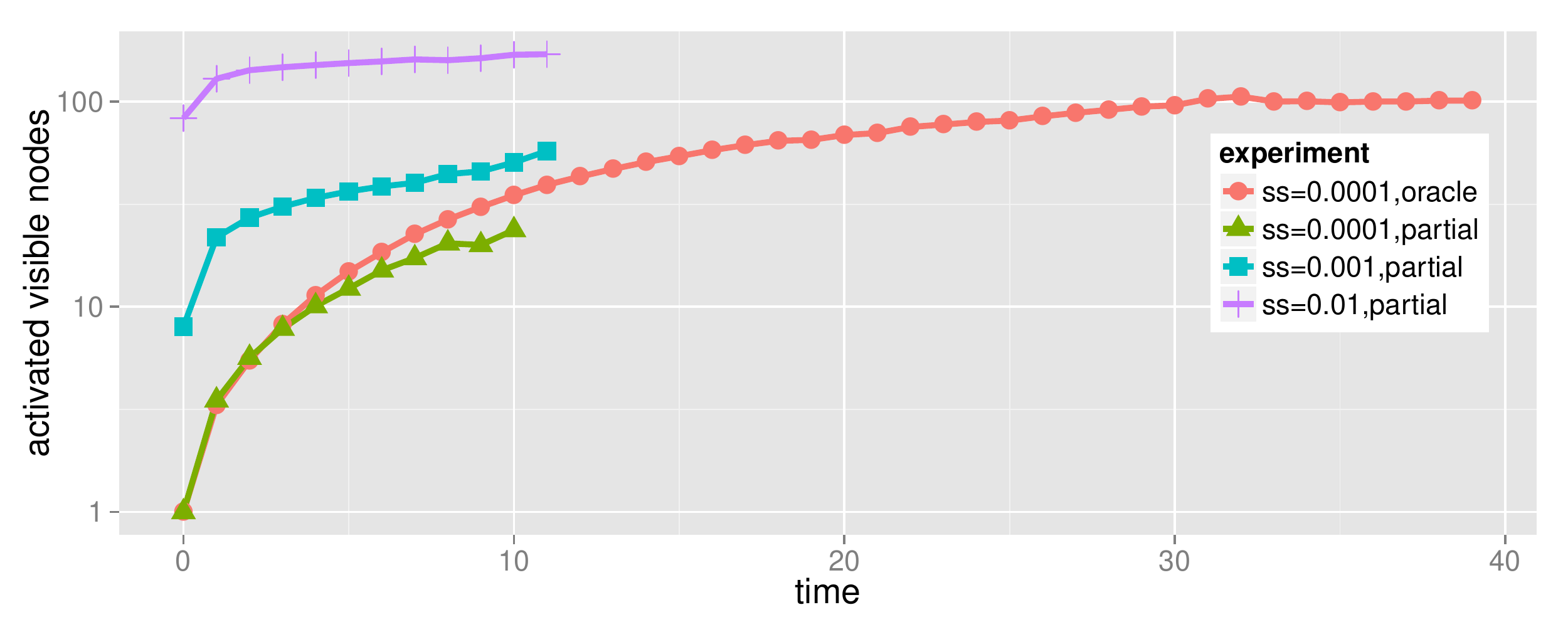}

    \caption{\small Fraction of the activated nodes at each time for different experiments for 2.5K.}
    \label{fig:delay}
  \end{center}

\end{figure}

Figure~\ref{fig:delay} presents the result of the diffusion process for both the partial and the oracle networks. The y-axis represents the average cumulative number of activated visible nodes\footnote{Note that since the means for later time steps were computed over a smaller number of cascades due to varying cascade lengths, the curves are not strictly monotonously increasing.}, and the x-axis represents time. 

We first examine the case where the number of initial users constitutes 0.01\% of the partial network (i.e., corresponding to only 1 user) to spread the campaign. For this seed size, the green curve (triangles) shows the diffusion launched over the partial network that is observed by the company, whereas the red curve (circles) represents the \emph{real} number of users that the campaign has reached at each time step (i.e., the oracle). The difference between the two curves demonstrates that the partial cascade size gives an inaccurate estimation of the total number affected users, showing that the campaign fails to reach the desired goal of 100 users. However, in reality there are many more users affected through the hidden portion of the network and the goal is reached at $t=30$.
 
Since the estimated cascade size based on the partial network appears to have failed to reach the campaign goal, the company may conceivably decide to allocate more resources by increasing the number of initial users. Figure~\ref{fig:delay} depicts  this scenario by showing the blue  (squares) and purple curve (pluses) which correspond to the seed size of 0.1\% (8 users) and 1\% (80 users), respectively. In the case of 0.1\%, the campaign again \emph{appears} to fall short of the desired number of users, while in the 1\% case the target is met almost immediately (although now 80\% of the goal are in fact initial users, making the investment substantially outweigh the return).
 
Allocating more resources to increase the seed size often incurs a high cost and may not be an option on large networks where an increase in the seed size may correspond to giving incentives to thousand of users.  As highlighted by this use case such actions may not always be necessary because the desired number of activated users would have been already reached in the oracle scenario.  However, as we point out in Section~\ref{sec:implications}, the increased seed size allows the campaign to reach the target much faster.
 
Through this example use case we demonstrated how the partial view of the network can impact how diffusion simulations can be perceived in practice and in turn negatively affect decision making. We now turn our attention to quantifying this estimation error and investigate the impact of the seed size and the hidden portion of the network in depth, before moving on to understanding the impact of network topology on cascade shape and spread.

\subsection{Magnitude of the problem}

\begin{figure*}
          \centering
        \begin{subfigure}[b]{0.5\textwidth}
\includegraphics[width=\textwidth]{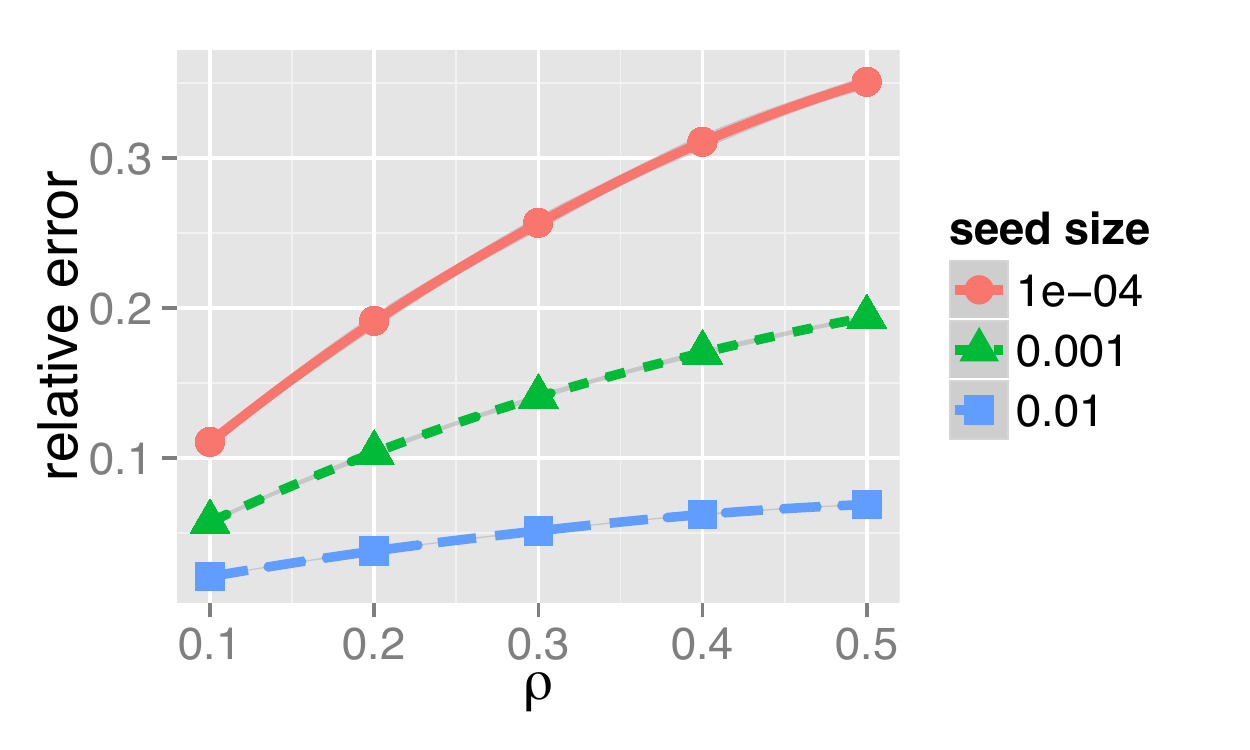}
                \caption{DBLP}
        \end{subfigure}%
        \begin{subfigure}[b]{0.5\textwidth}
           \includegraphics[width=\textwidth]{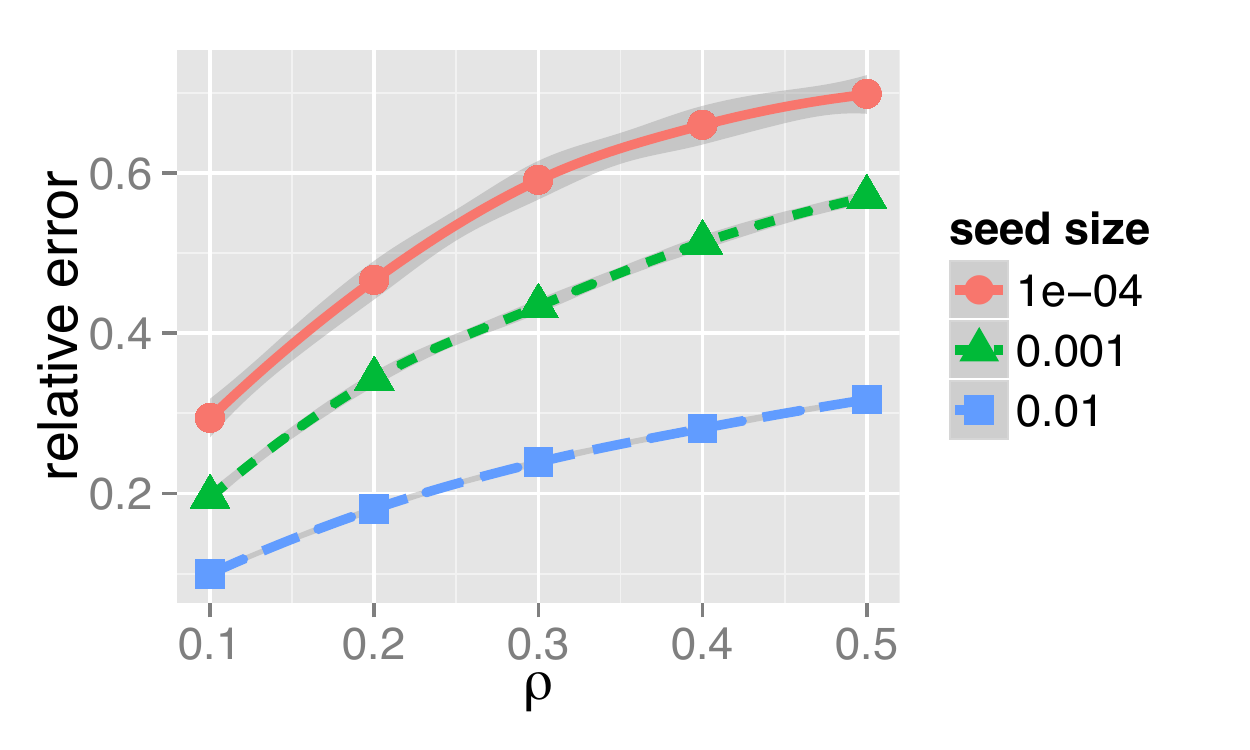}
                \caption{Astro Physics}
        \end{subfigure}
	\begin{subfigure}[b]{0.5\textwidth}
           \includegraphics[width=\textwidth]{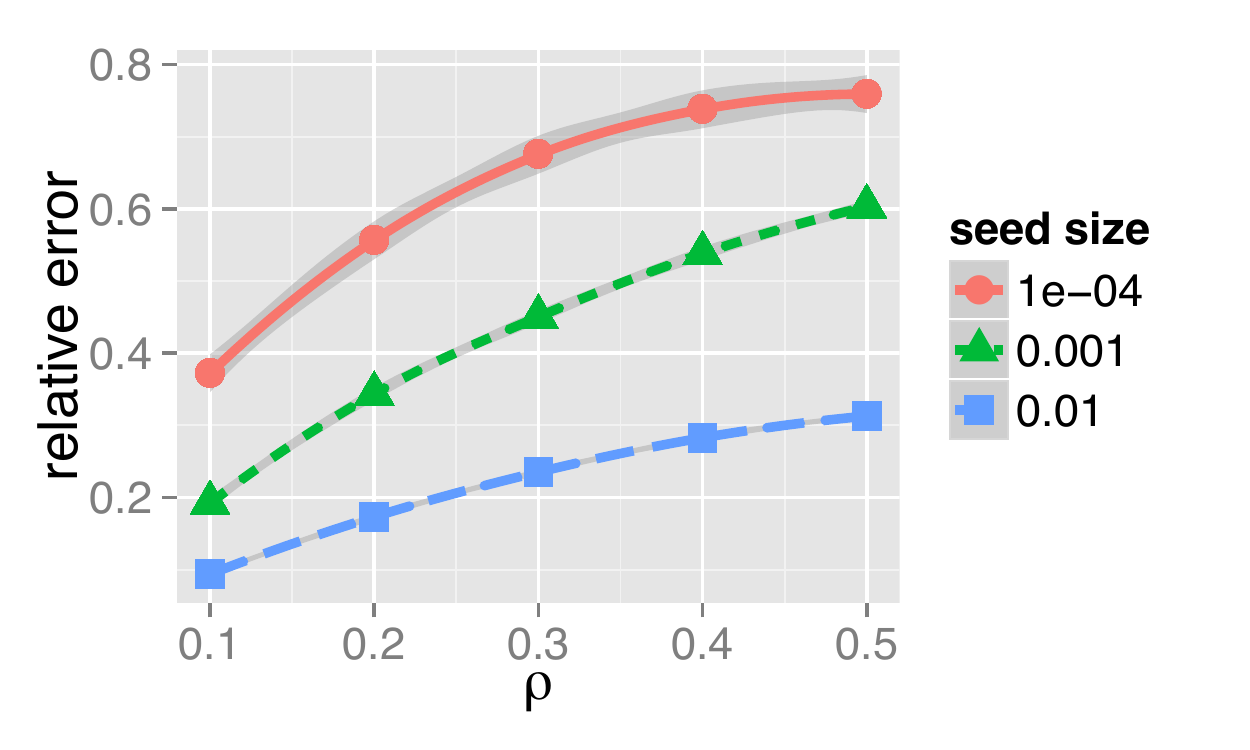}
                \caption{2.5K}
        \end{subfigure}%
         \begin{subfigure}[b]{0.5\textwidth}
   \includegraphics[width=\textwidth]{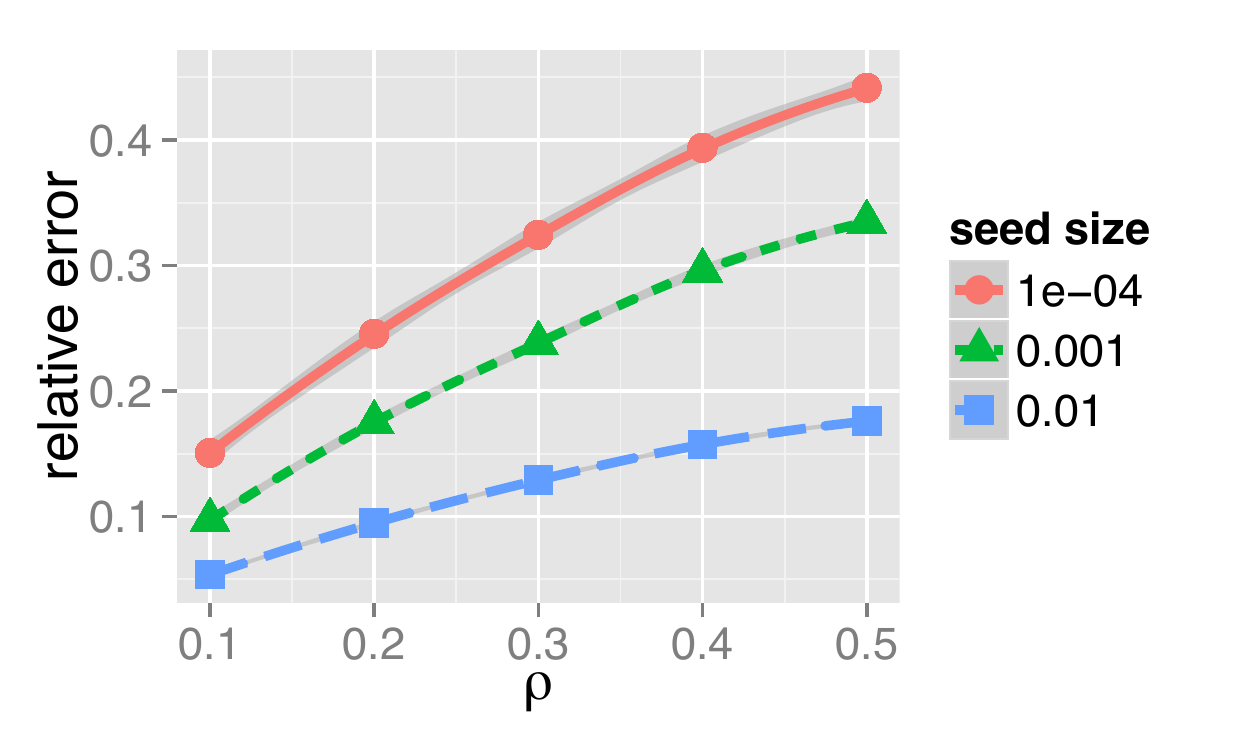}
                \caption{Toshk}
        \end{subfigure}
 \begin{subfigure}[b]{0.5\textwidth}

           \includegraphics[width=\textwidth]{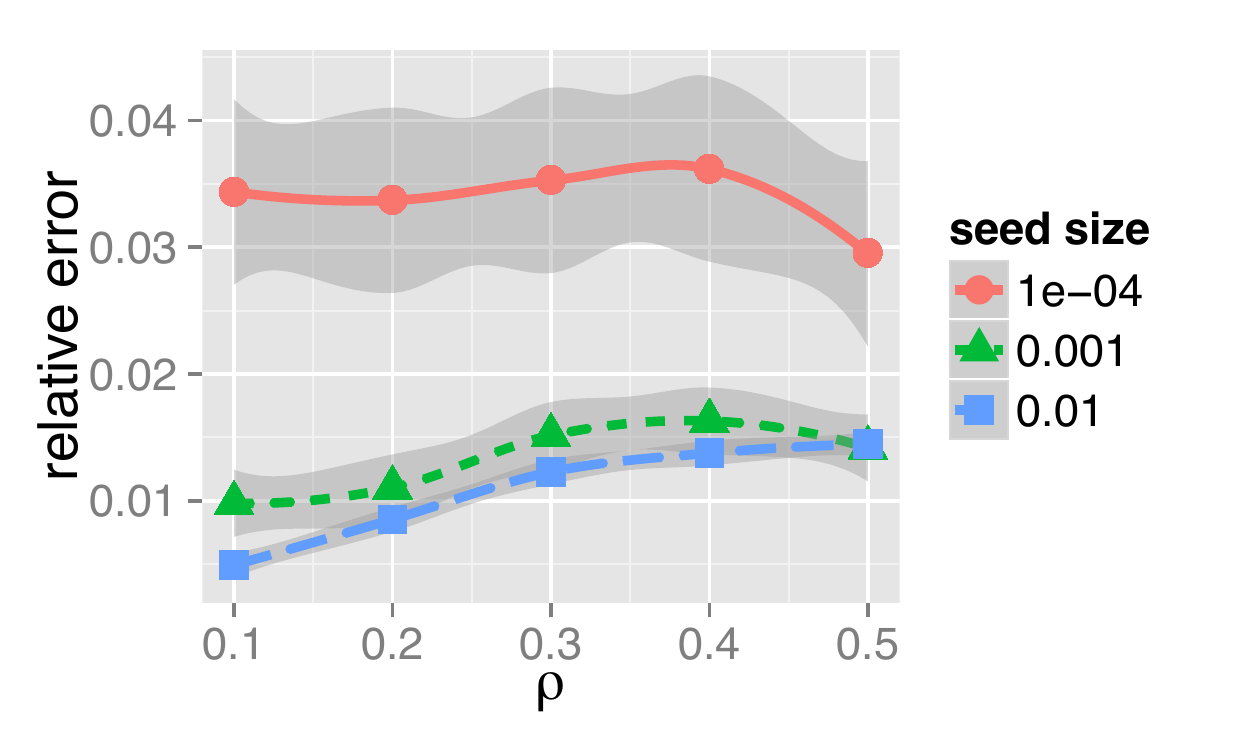}
                \caption{Erd\H{o}s-R\'{e}nyi }
        \end{subfigure}%
         \begin{subfigure}[b]{0.5\textwidth}
         \includegraphics[width=.2\textwidth]{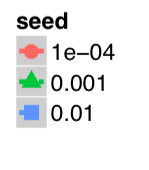}
         \vfill

          \end{subfigure}
   \caption{\small Relative Error for various seed sizes as a function of the fraction of hidden nodes $\rho$.}
    \label{fig:residuals}
\end{figure*}

As we have defined in Section \ref{sec:spreadresid}, we quantify the diffusion error in terms of relative error which measures the number of visible nodes theoretically activated with complete knowledge of network structure in relation to the number of nodes actually activated given partial knowledge that is typically available. Figure~\ref{fig:residuals} shows the relative error as a function of the fraction of hidden nodes $\rho$ for all networks. Each curve represents a different seed size $\gamma$ with 95\% confidence intervals. The relative error is considerably high for all networks and increases with an increase in $\rho$ and decrease in $\gamma$. Furthermore, this error is consistently due to \emph{underestimation}, as opposed to overestimation, of the number of the visible nodes in the partial network compared to the number of visible nodes in the oracle network and is a direct consequence of the hidden nodes. More specifically, in Eq. (\ref{eq:residual}) $(E(\sigma_{ph})+E(\sigma_o)) > E(\sigma_p)$. The exception to this is the Erd\H{o}s-R\'{e}nyi graph. Indeed, as we will show in the next section the topological structure (i.e., zero assortativity), yields uniform diffusion processes where a diffusion on the partial network can reach more visible nodes than on the oracle. Figure \ref{fig:residuals} shows evidence of this with the very low relative error and lower effect of the seed size. Because of this expected behaviour, we exclude it from further discussion in the following text.

 \para{Effect of the hidden nodes.} As more of the network is hidden (i.e., increasing $\rho$), the relative error also increases for all seed sizes. The curves begin to plateau for the higher values of $\rho$ because as the partial network gets smaller, the pool of initial visible nodes shrinks, meaning there are more hidden nodes which, when activated, are not accounted for in the calculation of Eq. (\ref{eq:residual}).

\begin{figure}[t]
\begin{center}
\includegraphics[width=\textwidth]{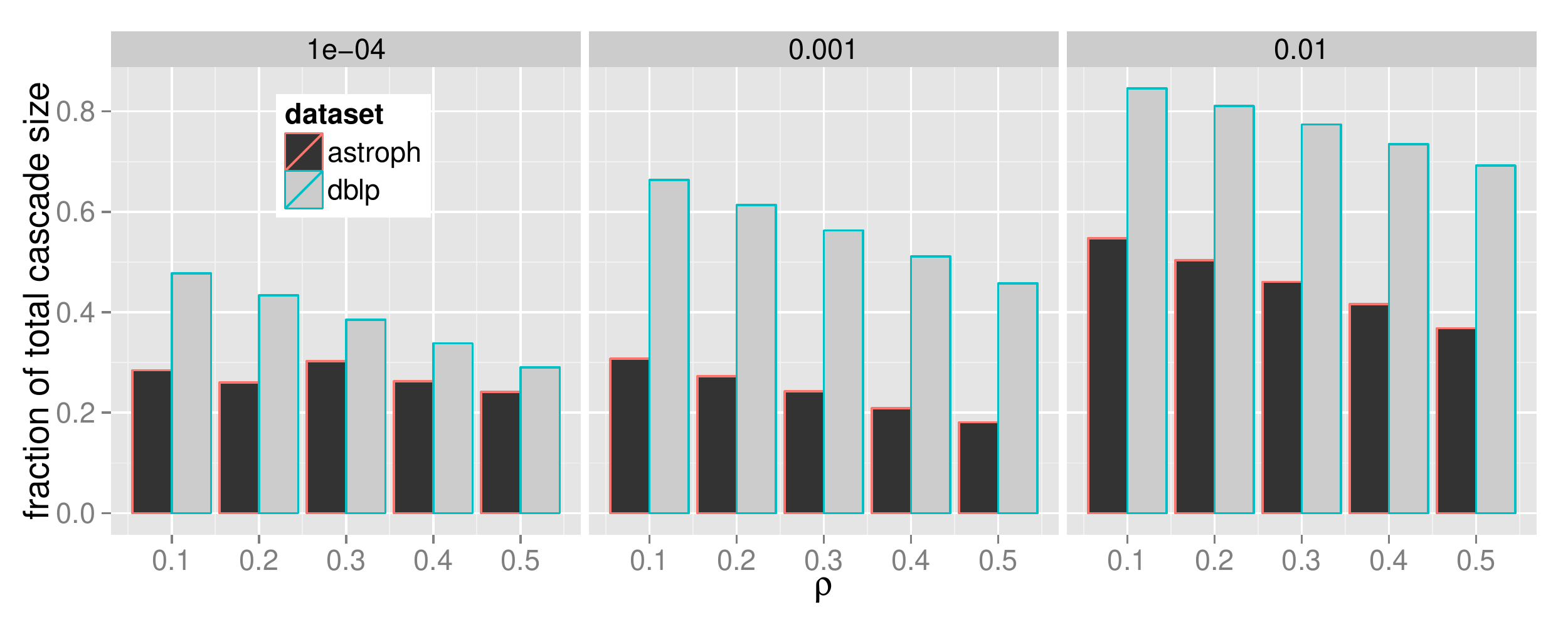}
 
\caption{\small Fraction of nodes that were observably activated at time 0 (seeds) and 1 (directly from the seeds) for the three seed sizes (columns) for the empirical networks.}
\label{fig:firstHopDiffusion}
\end{center}

\end{figure}

\para{Effect of seed size.} With the exception of Erd\H{o}s-R\'{e}nyi (as noted above), the error increases dramatically by reducing the seed size for all the networks, reaching approximately up to 0.8 in the worst case (2.5K with smallest seed and $\rho = 0.5$). The smallest seed size also results in a higher standard deviation, as would be expected due to the inherent stochasticity of ICM. This effect is most prominent for the smallest networks (Astro Physics and 2.5K) where the  seed fraction $0.0001$ corresponds to selecting just one or two seeds. Conversely, for the bigger networks such as DBLP (and TOSHK) the magnitude of the error is smaller for all the seed sizes and decreases as the number of seeds increases because the majority of the nodes are now activated  observely. Figure~\ref{fig:firstHopDiffusion} lends evidence to support this argument by depicting the  observed cascade size as a fraction of the total cascade size after the first iteration of ICM (i.e., the spread from the seeds to their direct neighbours), for DBLP and Astro Physics\footnote{TOSHK and 2.5K are omitted due to space constraints but share similar characteristics to DBLP and Astro Physics respectively}. In the case of the biggest seed size, up to 80\% of the final cascade is \emph{observably activated} as a result of direct activation from seeds, whereas for Astro Physics, this percentage is smaller, yielding higher chances of encountering hidden nodes later in the diffusion process, which in turn results in a higher relative error. In the next section, we discuss more on the shape and length of the cascades on different networks accounting for their topological structures.  

Finally, it is worth noting that while the relative error is decreased for larger seed sizes, increasing the seed size in real world scenarios is not always feasible. Indeed, as motivated by the use case detailed in the preceding text, the cost of targeting more users must be weighed against the perceived gain in the total cascade size. In most cases, simply increasing the seed size is prohibitively expensive, and in the absence of strict time constraints, may even be unnecessary. For these reasons, we shall concentrate on the smallest seed size ($\gamma=0.0001$) for the remainder of this section.

\subsection{Cascade Spread}
\input{inc/cascade-spread}

%% file: inc/cascade-spread.tex
The observation that the relative error in Eq. (\ref{eq:residual}) is dependent on the seed size, the fraction of hidden nodes and the number of first-hop (observable) activations leads to the following questions:

\vspace{-0.15cm}

\begin{description}
\item[i)] How does cascade tree shape affect the estimation error as a function of the fraction of hidden nodes?
\vspace{-0.25cm}
\item[ii)]How does the error relate to the topological properties of the graph?
\end{description}
\vspace{-0.15cm}

\begin{figure}[t]
\begin{center}

\includegraphics[width=\textwidth]{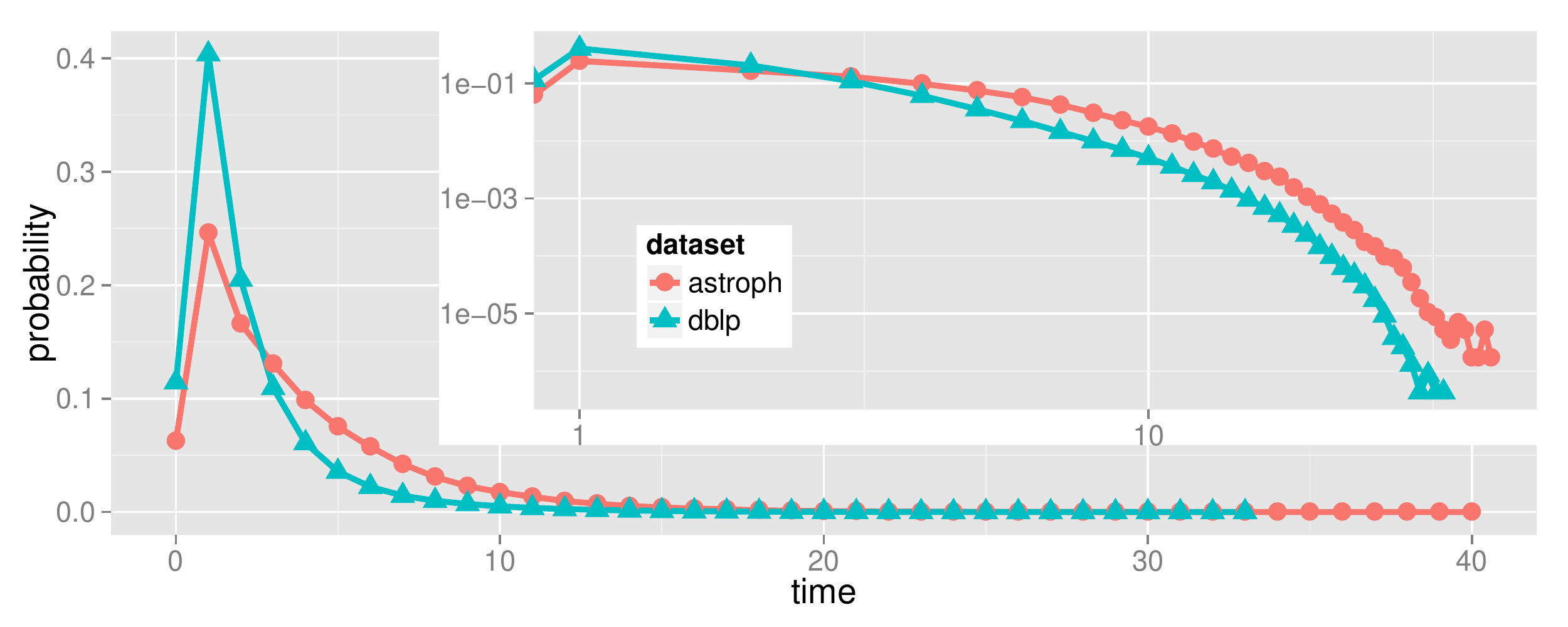}

\caption{\small Cascade shape for the empirical networks (inner plot is log-log).}
\label{fig:oracle_cascadedist}
\end{center}

\end{figure}

We answer the first question by showing that the trend that the relative error follows is a direct consequence of the cascade shape. Specifically, the branches of a wide and shallow cascade tree (and by extension, the {\un } cascades) will be small, even as more nodes are hidden by increasing $\rho$. Conversely, when a cascade tree has numerous long branches, increasing the number of hidden nodes will increase the size of the {\un } branches. In order to characterise the shape of these cascades we show the distribution of activated nodes at each time step. Figure~\ref{fig:oracle_cascadedist} shows that for the two empirical networks the cascades have different behaviour ($\rho=0.1$, $\gamma$=0.0001). Specifically, the cascades on DBLP activate a large number of nodes during the first few time steps, but new activations drop quickly in the subsequent time steps to finally end in relatively short branches. In contrast, the cascades over Astro Physics are smaller in the first few time steps, but maintain a higher number of newly activated nodes subsequent time steps, resulting in longer branches.

This explains why hiding more nodes, particularly those early on in the cascade, will result in larger consequent branches (i.e., {\un } cascades). The subplot of Figure~\ref{fig:oracle_cascadedist} shows the same empirical distributions on a log-log scale to emphasise the behaviour of the cascade tails. Astro Physics has a higher and longer tail which means that the likelihood of having many long branches is higher than for the other network. This also explains why the relative error in Astro Physics is double that of DBLP as reported in Figure~\ref{fig:residuals}.
\begin{table}[ht]
  \begin{center}
  \caption{\small Topological features of empirical (DBLP,  Astro Physics) and synthetic (TOSHK, Erd\H{o}s-R\'{e}nyi, 2.5K series) networks. Average degree (Deg), Average clustering coefficient (Clu), Diameter (Dia), Radius (Rad), Assortativity (Asso).}
  \label{tab:network-properties}
  \begin{tabular}{lccccc}
    \hline
    Network $(|V|,|E|)$                       &  Deg & Clu&  Dia& Rad& Asso\\
    \hline

    DBLP (1M, 4M)                        & 7.7 & 0.634      & 19$^\dagger$ & 12$^\dagger$ & 0.114	\\ 
    Astro Physics (18K, 198K)                 & 21.1 & 0.531     &14  & 8 & 0.205\\ 
     2.5K series (AstroPh) (16K, 158K)   & 19.0 & 0.552     & 10  & 6 & 0.223\\ 
    TOSHK (100K, 1M)                        & 21.2 & 0.528     & 11  & 6 & 0.043\\ 
    Erd\H{o}s-R\'{e}nyi (100K, 1M)         & 21.0 & 0.0002    & 6   & 5 & 0.0002 \\
    \hline
  \end{tabular}
  \end{center}
  \vspace{-0.25cm}
  \small
  \hfill$\dagger$ Estimated by sampling 1000 nodes averaged over 10 samples

\end{table}

We answer the second question by illustrating how clustering coefficient and assortativity play a fundamental role in affecting the shape of the cascades. 
Table~\ref{tab:network-properties} lists our selected networks and their most significant topological properties. These spectrum of
different networks ranges from purely random (Erd\H{o}s-R\'{e}nyi) with negligible clustering structure (also indicated by a low average clustering coefficient in Table~\ref{tab:network-properties}) to those indicating strong community structure, such as DBLP with an average clustering coefficient of $0.634$. The synthetic networks, with the exception of Erd\H{o}s-R\'{e}nyi, also have high average clustering coefficients because this topological metric is specifically targeted in the generation process. Table~\ref{tab:network-properties} also shows that real networks have positive assortativity indicating that nodes tend to associate with those of similar degree. The synthetic graphs show neutral assortativity which is consistent with many random graph models. The 2.5K series is the exception due to the explicit modelling of the JDD.

\begin{figure}

\includegraphics[width=\textwidth]{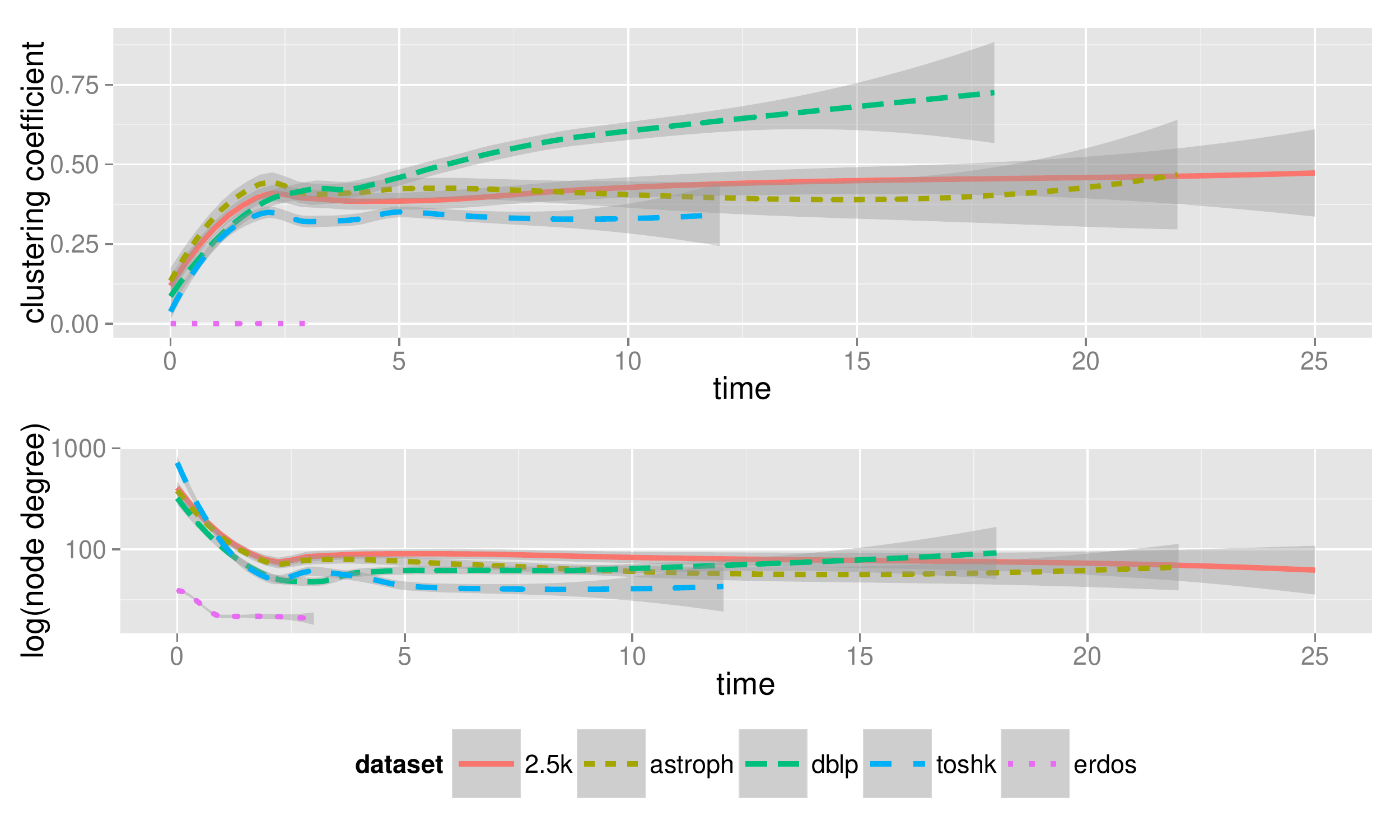}

\caption{\small Average node clustering coefficient and degree as a function of node activation depth in the partial cascades for the empirical networks.}
\label{fig:clustering_by_level}
\vspace{-1em}
\end{figure}

 Figure~\ref{fig:clustering_by_level} (upper) shows the average clustering coefficient for nodes activated at each time step ($\rho=0.1$, $\gamma=0.0001$) with shaded areas as 95\% confidence intervals. The average clustering coefficient for DBLP continually increases as a function of time, while for the other networks it remains low and relatively constant after the second time step. Highly clustered and disassortative nodes have a lower chance of spreading the information widely for two fundamental reasons. First, once the diffusion enters a clique, it will likely remain within that portion of the graph~\cite{ver2011stops}. Second, recall from Table~\ref{tab:network-properties} the lower assortativity of DBLP (0.114). Low degree heterogeneity has been shown to limit cascade size \cite{ver2011stops} because spreading to high-degree nodes is less likely, especially when seeds are chosen with discounted degree. This last point can be confirmed by examining average node degree at each time step.

In Figure~\ref{fig:clustering_by_level} (lower), we consistently observe across all networks that the majority of high-degree nodes are activated during the initial time steps. The networks that activate nodes with lower degree, e.g., TOSHK and Erd\H{o}s-R\'{e}nyi, also tend to have shorter cascade branches which in turn leads to lower probability of nodes being activated through hidden nodes (hiddenly activated) as the cascade progresses.

%% file: inc/results_theo.tex
\section{Correction}
\label{sec:prediction}
As we showed in the previous section, the {\un } cascade can have a dramatic effect on the estimates of the number of activated visible nodes. Therefore, in this section we focus on the problem of predicting the total cascade size given the partial network, the fraction of hidden nodes $\rho$, and seed size $\gamma$. Having knowledge of $\rho$ is a reasonable assumption in cases when we know the total population, and therefore we can easily estimate how many nodes are missing in the network. For instance, in the case of the telecommunication operator or social media campaigner, the total population may be known from a census or can be estimated by a poll. We experimented with two approaches to the correction of the partial cascade size. Both methods use the partial cascade size $\sigma_p$ assuming that it closely approximates a part of the total cascade and then attempt to correct it by adding the remainder of the cascade. Although we are aware that error correction for synthetic networks does not have much practical value, we still investigate the results for synthetic networks in order to test the robustness of the presented methods.

\begin{figure}
        \centering
        \begin{subfigure}[b]{0.45\textwidth}
                \centering
                \includegraphics[width=\textwidth]{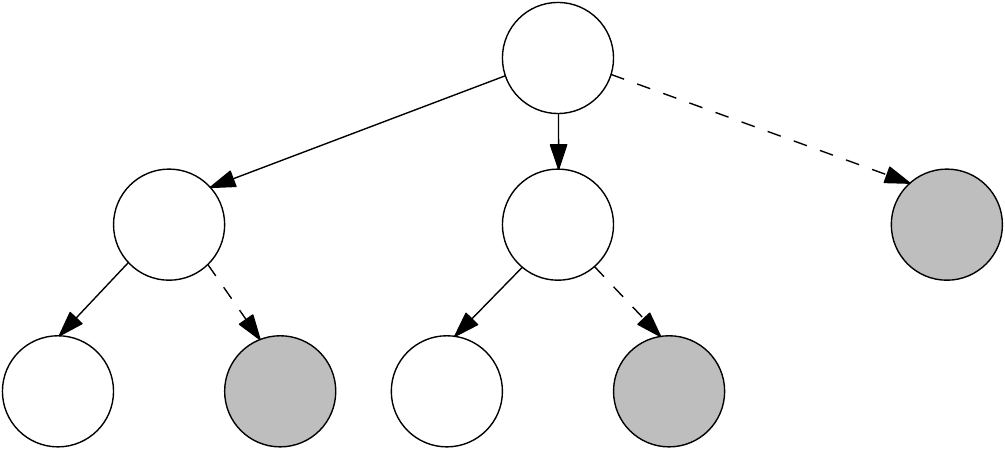}
                \caption{SiCE}
                \label{fig:sice}
        \end{subfigure}%
        ~ 

        \begin{subfigure}[b]{0.45\textwidth}
                \centering
                \includegraphics[width=\textwidth]{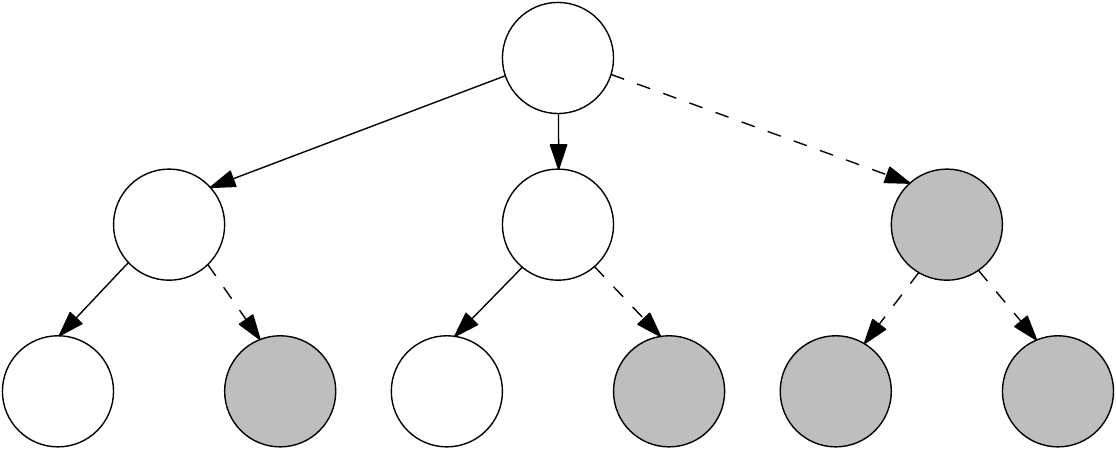}
                \caption{ReCE}
                \label{fig:rece}
        \end{subfigure}
        \caption{The two approaches to correction of partial cascades. The visible nodes that were activated during the
          diffusion over the partial network are white and the grey nodes represent the unexplored nodes (in the partial cascade) added by the
          correction method. The solid links represent the activation pathways as observed during the simulation, while
          the dashed links represent the additional activations of the unexplored nodes induced by the correction method.}\label{fig:correction}
\end{figure}

\para{Simple Cascade Expansion (SiCE)} The SiCE approach realizes a simple correction of the partial cascade by inferring the size of the unexplored nodes as proportional to the $\rho$ fraction of missing nodes from the partial network.

This approach assumes that all nodes activated in the partial network, i.e.
$\sigma_p$, were also the visible nodes activated in the oracle network, thus $\sigma_p$ is assumed to be a tight approximation of $\sigma_{ph}+\sigma_o$.

Hence, the total cascade size $\sigma$, in the oracle network, can be approximated as $\hat{\sigma}=\sigma_p/(1-\rho)$, which corrects the partial cascade size by adding the expected number of unexplored hidden nodes. This corresponds to estimating the expected number of descendants, for each node $i$ in the partial cascade, that would have been unexplored compared to the oracle cascade. Let $d_p(i)$ be the number of $i$'s direct descendants in partial cascade, then $d_p(i)/(1-\rho)$ would be the expected number of $i$'s direct descendants in oracle cascade; thus, to correct the partial cascade we only need to add, for each node i, the respectively missed nodes, i.e. $d_p(i)/(1-\rho)-d_p(i)$, as we show in gray colour in Figure~\ref{fig:sice}.

\para{Recursive Cascade Expansion (ReCE)} So far, we have shown that the oracle and partial cascades often differ
significantly (recall Figure~\ref{fig:residuals}) due to the large {\un } cascade, therefore, the assumption $\sigma_p\approx \sigma_{ph}+\sigma_o$ in the SiCE approach does not strictly capture reality. Here, we introduce a different approach, i.e. ReCE, in which we assume that the partial cascade size $\sigma_p$ is a good approximation only of the {\em observed} portion, 
$\sigma_o$, of the oracle cascade. Thus, ReCE tries to infer the total cascade size
by growing the partial cascade with new nodes while letting them diffuse further to account for the  {\un } portion of the oracle cascade. As depicted in Figure~\ref{fig:rece}, ReCE expands the cascade with a top-down approach level by level by inferring at each level $t$, the expected total cascade $\hat{\sigma}^t$. ReCE starts from the nodes directly activated
by seeds, i.e. at level 1, where it is equivalent to SiCE. 
On the subsequent levels, it adds also the estimated number of descendants of the new 
nodes from the previous level, i.e. $\hat{\sigma}^{t-1}-\sigma_p^{t-1}$, as computed in the following formula: 
\begin{equation}
  \label{eq:1}
  \hat{\sigma}^t=
  \begin{cases}
    \frac{\sigma_p^t}{1-\rho}+(\hat{\sigma}^{t-1}-\sigma_p^{t-1})E(deg^{t-1})p & \quad \text{for } t\ge 2\\
    \frac{\sigma_p^t}{1-\rho} & \quad \text{otherwise}
  \end{cases}
\end{equation}
where $E(d^{t-1})$ is the expected number of the visible descendants of a node activated at
level $t-1$ and $p$ is the transmission probability. This method is expected to outperform the simple correction model, i.e. SiCE, because it allows the new added nodes to spread the diffusion process further. 
\begin{figure}[t]
   \begin{center}
     \includegraphics[width=\textwidth]{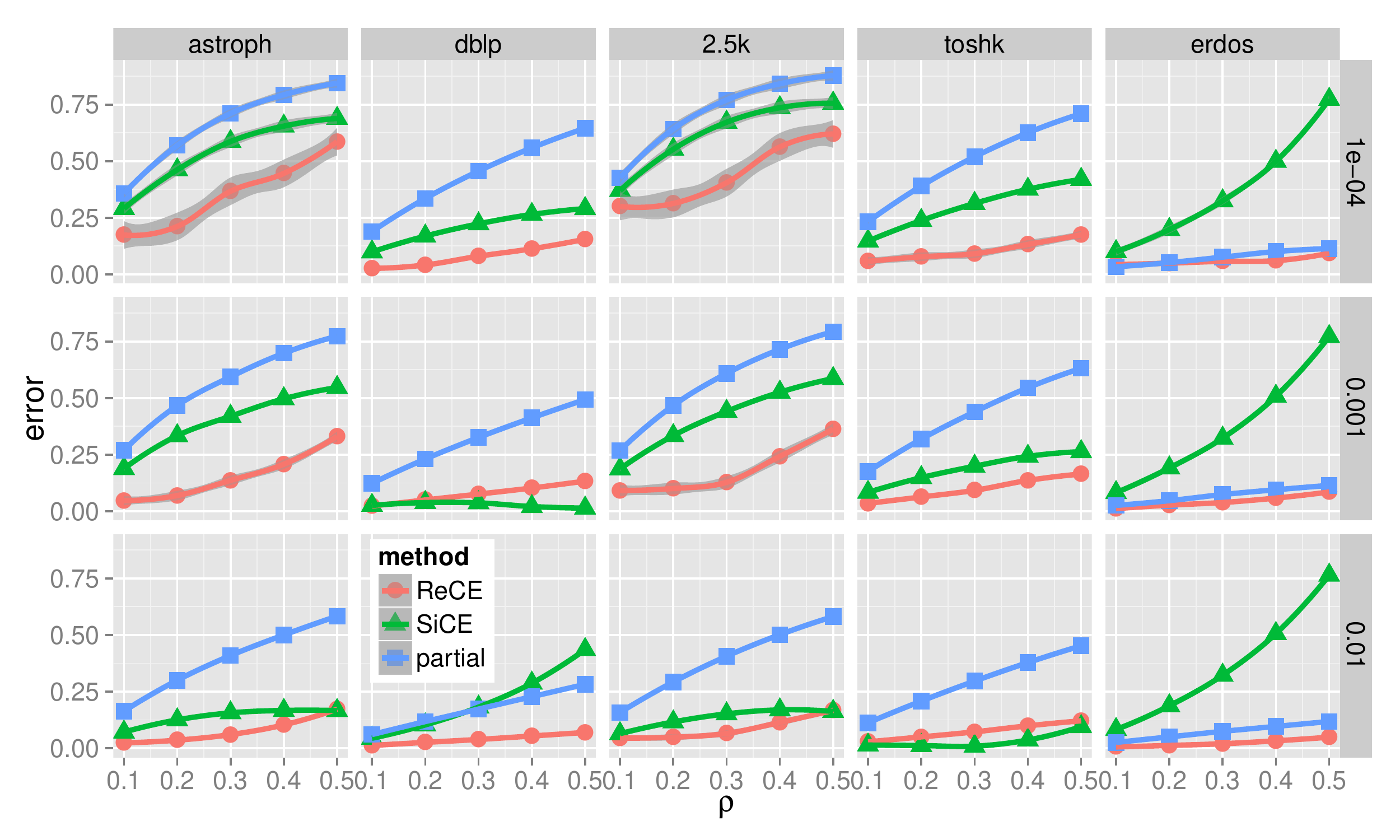}
     \caption{\small The mean absolute relative error of three approaches to correction of partial cascade size for the networks (columns) and seed sizes (rows).}
     \label{fig:prediction}
   \end{center}
 \end{figure}
In the next paragraph, we compare this two approaches together with the non-corrected partial cascade by measuring their mean absolute relative error  with the oracle cascade, i.e. 
$E(|\sigma-\hat{\sigma}|/\sigma)$, recall that $\sigma$ is the total cascade size. 

 \para{Model validation.} Figure~\ref{fig:prediction} shows the relative error of three approaches in predicting of the
 total cascade size $\sigma$. The mean values are denoted by points which are interpolated by a smoothed line with a shaded 95\%
 confidence interval. The \emph{partial} approach uses the partial cascade size without any correction. In general we observe that both SiCE and ReCE feature smaller error than the partial method, and ReCE provides the overall lower error among all. Note that the Erd\H{o}s-R\'{e}nyi
 network is an exception in the sense that the error induced by SiCE was the highest of all the three methods. On Erd\H{o}s-R\'{e}nyi the cascades are particularly shallow, therefore blind correction, as for SiCE, quickly
 overestimates the cascade size. Conversely, ReCE, which expands the cascade layer by layer, seems to be more robust and induce consistent results across all the networks.
 We identify two cases in which SiCE slightly outperforms ReCE (on the TOSHK network with the highest seed size, and on the DBLP network with the middle seed size). This observation highlights that when the cascades start from many seeds and the visible part of the networks is quickly covered, sophisticated correction methods do not provide substantial benefits because the {\un } cascade size is negligible.

Generally, ReCE led to the smallest correction errors, in some cases improving the estimates of the oracle cascade size
 notably, e.g., for DBLP, smallest seed size, and 50\% missing nodes ReCE induced 15\% error, reducing the error by 49\%
 (compared with the partial method). The relatively high error and broader confidence interval for Astro
 Physics and the smallest $\gamma$ is caused by the very small number of seeds, and we see that the error decreases
 dramatically for the medium seed size. However, as $\rho$ increases, the
 difference between the distributions of activation times on the oracle and partial networks grows which in turn
 increases the error.

%% file: inc/implication.tex
In this section we first present several theoretical and practical implications of this work. After that, we discuss a few limitations of our approach and suggest ways to address them.

\subsection{Implications}
\label{sec:implications}
\para{Theoretical Implications.} This study has quantified cascade size estimation error caused by network partiality and has offered a way to correct for it. In particular, this work highlights that the cascades can be much deeper than what they appear to be in the partial network. This observation has also been confirmed by \cite{dow2013anatomy} on the \emph{real} cascades based on photo sharing on Facebook, where the information that previously appeared to have been shared directly  from the source was traced back to a considerably greater depth in the cascade. We believe this work has direct implication on how the diffusion process is perceived and explained by the research community.

\para{Practical Implications.} Understanding  \emph{`how many'} nodes have been activated  and \emph{`how'} they were reached can be leveraged by campaign managers and marketing companies to assist them with decision making. Based on the results presented in this work,  we argue that if one requires the campaign to spread quickly (e.g., a time-sensitive petition), then more resources should be allocated to engage a larger set of initial users.  On the other hand, for a word-of-mouth campaign which is delay tolerant and does not require fast spread (e.g., advertisement), the desired number of users can be still reached in a longer period of time through the hidden parts of the network. Finally, our correction method enables those relying on diffusion simulations to get a more accurate understanding of the of the magnitude that the campaign has actually reached assisting them to allocate their resources more efficiently.

\subsection{Limitations}

This work has a number of limitations. First, we modelled the diffusion process based on the ICM. While we believe the problem addressed in this paper still exists and can be quantified in the same way for other diffusion algorithms such as the linear threshold model (LTM)~\cite{kempe2003maximizing}, the correction process will differ significantly.  

Second, we confined our experiments to the most common case of global transmission probability $p=0.01$, which has been adopted widely by the literature~\cite{kempe2003maximizing,chen2009efficient} and has been empirically measured in online social networks~\cite{steeg2011stops,Bakshy:2011,Bakshy:2012:RSN:2187836.2187907}. If we were to try with higher probability, we would observe that the diffusion process reaches the saturation point quickly~\cite{chen2009efficient}, resulting in short and wide cascades with small relative error between the oracle and the partial cascade sizes.

A global transmission probability assumes that all nodes have the same threshold for diffusion. While this may hold in some cases such as with models of disease spread, it is simplistic to assume that it applies to real world social networks where users are influenced by others in more nuanced ways. Modelling varied social tie strengths instead of a uniform global transmission probability is straightforward: ICM has already been extended to weighted edges \cite{Kempe2003}. In this extension, a biased coin flip according to the edge weight replaces the uniform transmission probability at each edge. With respect to our work, the cascade size measurement remains unchanged: if the activated node is hidden, then it is attributed to the hidden cascade; if it is visible (including the entire sub-tree up to the seed), then it is attributed to the observed cascade. If neither are true, it is attributed to the phantom cascade.

Correcting the relative error, on the other hand, is slightly trickier. We can recursively expand the cascade size using the average degree (at the cascade level) and the transmission probability (Eq. \ref{eq:1}). For ReCE, we posit that for the size of the visible cascade at each level, we could simulate over the weighted degree distribution one layer above (this would replace the expected degree at the previous level $\times p$). In other words, we cannot take the product of the average degree and $p$, since the weights would differ, so we would have to take the weighted degree distribution instead. The other correction method we have introduced, SiCE, does not require any modification for adaptation to weighted edges.
 
Finally, we only reported the results of our experiments where the seed nodes were selected by their discounted degree, but we also conducted experiments with seeds selected uniformly at random to control for a possible bias introduced by the seed selection strategy. Although the variance of the measured error metrics rose, our findings were consistent with the results that are presented.

%% file: inc/conclusion.tex
\section{Conclusion}\label{sec:conclusion}
\label{sec:conclusion}

In this paper, we focused our investigation on the impact of network partiality on information diffusion processes and
have demonstrated the magnitude of the estimation error when only partial data is available.  Our work provides a novel
methodology to characterise this error and sheds light on features that allow us to correct for it.

While we have taken the first step toward the correction of the estimated error based on partial data, we believe our results can be extended to account for different classes of graphs with different topological  properties in other domains. Also another valuable extension to our work would  be obtaining a more pragmatic way (e.g., community based) of sampling the hidden part of the network that would correspond to  individuals preferences in real world networks such as mobile networks. As part of the same stream of future work, we are also interested to investigate \emph{`who'} are the nodes that are activated and focus more on their topological characteristics (e.g., degree, betweenness centrality, their community etc.) rather than their sheer magnitude.